\documentclass[review]{elsarticle}

\usepackage{lineno,hyperref}
\journal{Journal of \LaTeX\ Templates}

\biboptions{authoryear}

\begin{document}
\newcommand{\adv}{    {\it Adv. Space Res.}}
\newcommand{\annG}{   {\it Annales Geophysicae}}
\newcommand{\aap}{    {\it Astron. Astrophys.}}
\newcommand{\aaps}{   {\it Astron. Astrophys. Suppl.}}
\newcommand{\aapr}{   {\it Astron. Astrophys. Rev.}}
\newcommand{\ag}{     {\it Ann. Geophys.}}
\newcommand{\aj}{     {\it Astron. J.}}
\newcommand{\apj}{    {\it Astrophys. J.}}
\newcommand{\apjs}{    {\it Astrophys. J. Suppl.}}
\newcommand{\apjl}{    {\it Astrophys. J. Lett.}}
\newcommand{\apss}{   {\it Astrophys. Space Sci.}}
\newcommand{\bain}{   {\it Bulletin of the Astron. Inst. of the Netherlands}}
\newcommand{\cjaa}{   {\it Chin. J. Astron. Astrophys.}}
\newcommand{\gafd}{   {\it Geophys. Astrophys. Fluid Dyn.}}
\newcommand{\grl}{    {\it Geophys. Res. Lett.}}
\newcommand{\ijga}{   {\it Int. J. Geomag. Aeron.}}
\newcommand{\jastp}{  {\it J. Atmos. Solar Terr. Phys.}}
\newcommand{\jgr}{    {\it J. Geophys. Res.}}
\newcommand{\mnras}{  {\it Mon. Not. Roy. Astron. Soc.}}
\newcommand{\nat}{    {\it Nature}}
\newcommand{\pasp}{   {\it Pub. Astron. Soc. Pac.}}
\newcommand{\pasj}{   {\it Pub. Astron. Soc. Japan}}
\newcommand{\pre}{    {\it Phys. Rev. E}}
\newcommand{\solphys}{{\it Solar Phys.}}
\newcommand{\sovast}{ {\it Sov. Astron.}}
\newcommand{\ssr}{    {\it Space Sci. Rev.}}
\begin{frontmatter}

\title{Extreme solar storms based on solar magnetic field}
%\tnotetext[mytitlenote]{Fully documented templates are available in the elsarticle package on \href{http://www.ctan.org/tex-archive/macros/latex/contrib/elsarticle}{CTAN}.}

%% Group authors per affiliation:
\author{Brigitte Schmieder \fnref{myfootnote}}
\address{Observatoire de Paris, LESIA, 5 place Janssen, 92195 Meudon, France.}
%\fntext[myfootnote]{}

%% or include affiliations in footnotes:
%\author[mymainaddress,mysecondaryaddress]{Elsevier Inc}
%\ead[url]{www.elsevier.com}

%\author[mysecondaryaddress]{ }
\cortext[mycorrespondingauthor]{Corresponding author}
\ead{brigitte.schmieder@obspm.fr}

\address[mymainaddress]{PSL Research University, CNRS, Sorbonne Universit\'es, UPMC Univ. Paris 06}
\address[mysecondaryaddress]{Univ. Paris Diderot, Sorbonne Paris Cit\'e}

\begin{abstract}

Many questions have to be answered before understanding the relationship between the emerging magnetic flux through the solar surface and the extreme geoeffective events.
Which threshold determines the  onset of the eruption? What is the upper limit  in energy for    a flare?  Is the size of sunspot the only criteria to get extreme  solar events?

Based on observations  of previous solar cycles, and theory,  the main ingredients for getting X ray class   flares and large Interplanetary Corona Mass Ejections e.g.  the built up of the electric current in the corona,  are  presented such as the existence of  magnetic  free energy,  magnetic helicity, twist and stress in active regions.
The upper limit of solar flare energy in space research  era and the possible chances to get  super-flares  and extreme solar events  can  be predicted  using MHD simulation of coronal mass ejections.

\end{abstract}


\end{frontmatter}

%\linenumbers
\section{Introduction}

%Starting point of our study which aims to trace back the ICME and identify the initial coronal mass ejection (CME) emitted at the Sun \citep{Tousey1973}. \\

\begin{figure}
 \centering
  \mbox{
   \includegraphics[width=12cm]{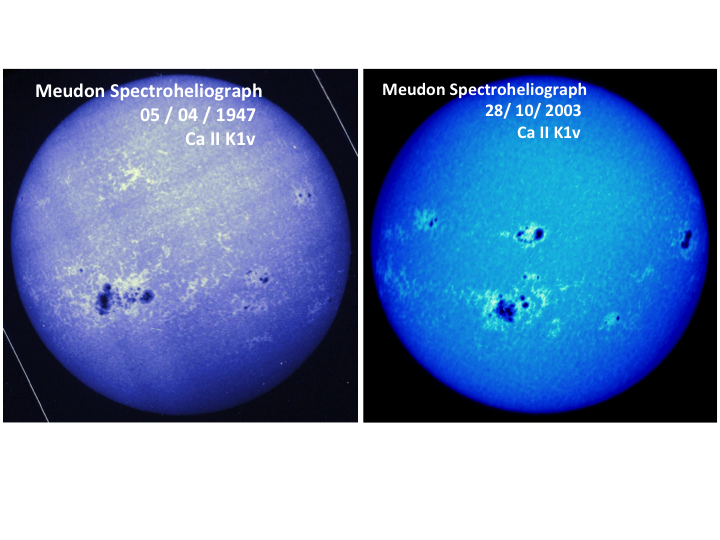}
  }
 \hspace{0.5cm} 
 \mbox{
  \includegraphics[width=12cm]{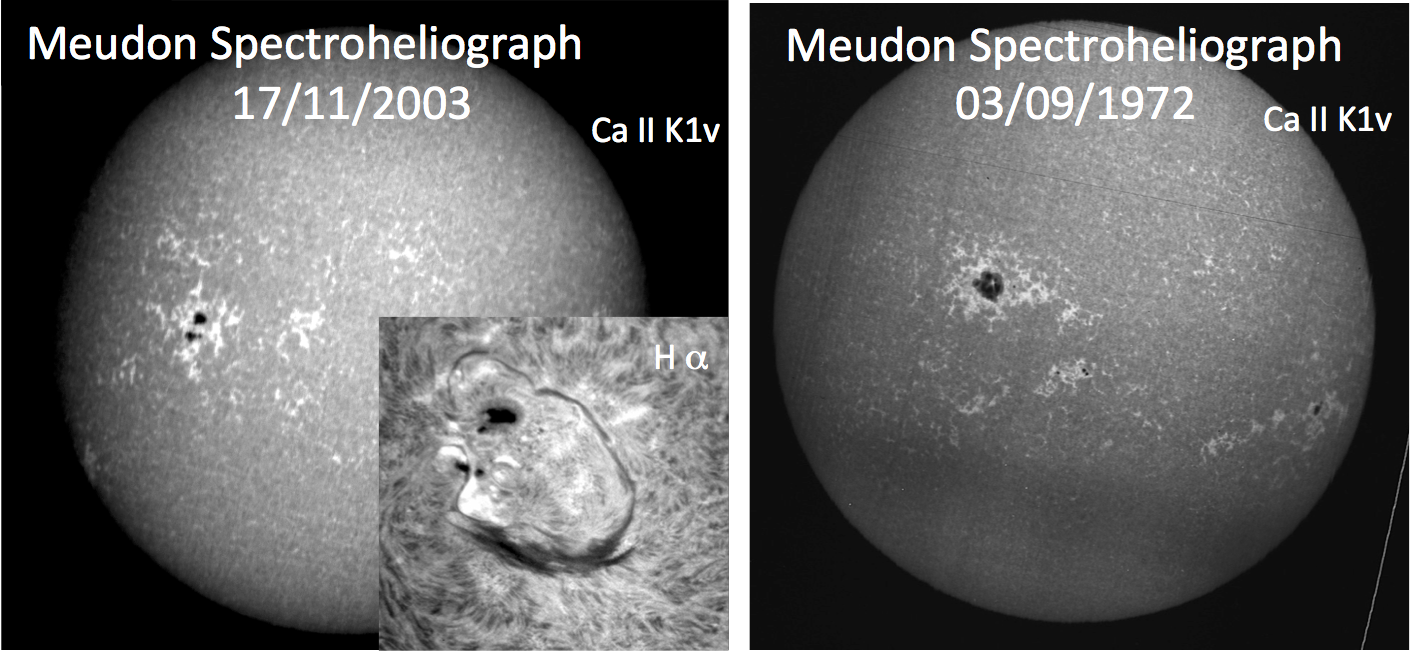}
  }
  \hspace{-5 cm} 
\caption{Full disk spectroheliograms from  the Observatoire de Paris in Meudon. ({\it top panels}) The  largest sunspot  groups ever reported:  ({\it left}) on April 4, 1947 with no geoffective effect,   ({\it right}) on October  28  2003. The    AR 10486 in the south hemisphere { led}  to a X17 flare  and  consequently a geomagnetic disturbance with   a Dst=-350 nT.({\it bottom panels}): %Large sunspot in October 2014 site of   a X class   flare and no CME (adapted from \citet{Sun15}), 
({\it left})  AR  10501 on November 17  2003 observed in Ca II KIv  with an inserted  H$\alpha$ image of the active region.   The  huge eruptive filament  surrounding the AR initiated 
%  and  mixed magnetic { polarity}  leading to a M9.6 flare and
 the largest Dst of the 23$^{th}$ solar cycle  (Dst=-427 nT). ({\it right})  McMath region 11976  large sunspot,  source of flares and  ejected energetic particles on August 1972
 (spectroheliograms from  the Meudon data-base ''BASS2000'').  }
\label{spot}
\end{figure}

Extreme solar storms can be defined as energetic solar events  related to  
large-scale disturbances in the Earth's magnetosphere, called  as geomagnetic events \citep{Cliver04,Koskinen06, Echer11a,Echer11b,Echer13,Gonzalez11b}.
{ Before the launch of satellites, the activity of the Sun was recorded by ground-based instruments observing in visible light 
(e.g. see the Meudon data-base ''BASS2000'' with  spectroheliograms  registered from 1909 until today- see examples in Figure \ref{spot}). Surveys in white light,  in H$\alpha$, and  Ca II H and K  lines allow to study the  solar cycle activity by tracking the sunspots and  studying  their size, and their complexity \citep{Waldmeier1955,McIntosh1990,Eren17}. The enhancement of  emission was  used as  a good proxy for detecting  flares \citep{Carrington1859}. However  the detection of flares was limited by the spatial and the temporal resolution of the observations.}

% was discussed during the International Geophysicalyear in 1957. Before that year, the geomagnetism was defined by local magnetic disturbances registered by magnetometers.The geomagnetic storms are characterized by solar-wind magnetosphere energy coupling enhancement and the growth of ring current on a short-time scale\citep{Saiz13}.  They are classified by the intensity of the $Dst$ as  weak ($Dst > -50$ nT), moderate  ($-100 <  Dst < -50$ nT), intense ($-200 <  Dst < -100$ nT),big  ($Dst <-200$ nT), and extreme storms  ($Dst  < -400$ nT) \citep{Koskinen06, Echer11a,Echer11b,Echer13,Gonzalez11b}.

Recently different approaches have succeeded to quantify the intensity of some historical events using different  magnetometer stations over the world.
The analysis of magnetic recordings made as early as the middle of the nineteenth century by %Indian % correction by AUDE
ground stations allowed us to clarify the importance of several extreme events {  \citep{Tsurutani2003,Cliver04,Lakhina2008,Cid13,Cliver2013}.}
During the XX$^{th}$ century, several important events with $Dst  < -700$ nT were observed after intense flares and connected to aurora. 
Exploring historical extreme events shows all the problems encountered when one aims at understanding the phenomena from one end to the other.
{ 	It  is difficult to identify  the solar source of extreme geoeffective events without continuous observations of the Sun and without quantified  numbers of the energy release during the solar events.	

The  {\it Geostationary Operational Environmental Satellites} (GOES)  register the global soft X ray emission 1- 8 \AA\, of the Sun since the ''80s''. The intensity of the flares are   classified by the letters X, M, C,  which correspond to 10$^{-4}$,  10$^{-5}$, 10$^{-6}$ W m$^{-2}$ energy release respectively. The extreme historical solar events, for which only the size of  sunspots and "the magnetic crochet" recorded on the Greenwich magnetogram, for example,  for  the Carrington event or  ionospheric disturbances   are known,  were associated with extreme geomagnetic events  by comparison  with  recent events. 
%The Carrington event could have correspond to a X ray  flare  of class $>$ X10
It is interesting to   read the papers of  \citet{Tsurutani2003,Cliver2013}  where  several historical events e.g. Sept. 1859, Oct. 1847,  Sept.1849, May 1921 have been discussed and classified.}

With the {\it SOlar and Heliospheric Observatory }\citep[SOHO;][]{Fleck1995},  launched in 1995, and its on-board spectro-imagers and coronagraphs, and more recently with the {\it Solar TErrestrial RElations Observatory}  {  \citep[STEREO  A and B 2006;][]{Wuelser2004,Russel2008} } and its  { COR}  and { HI}  coronagraphs able to reach the Earth in particular conjunction  (see the website of HI 
HELCATS)
%\verb{http:helcats.UK )
  the solar sources of   geoeffective events could be identified with more accuracy. A new era was open for  { forcasting  geomagnetic disturbances by   being able to follow   the solar events  in multi-wavelengths, and  particularly  the coronal mass ejections from the Sun to the Earth. This is the new science called  "Space Weather''.}
Intense flares responsible for geoeffective events are commonly associated with  Solar Energetic Particles (SEP) events and/or coronal mass ejections (CMEs). Several minutes after the flares, very high energetic particles (SEPs) may enter in  the Earth's atmosphere affecting astronauts or electronics parts in satellites. 
%{ They are registered by the neutron monitors (Turner,Klein).}
However, concerning geomagnetic disturbances, CMEs { can be as  geoeffective as  the energetic particles when their arrival trajectory is oriented towards the Earth and when their speed is large enough  \citep{Gopalswamy10a,Gopalswamy10b,Wimmer14}.  SEP ejections produce particle radiation }with large fluence,  however only a few of SEPs occur during each  solar cycle while  CMEs have an occurrence  rate between { 2 and  3 per week in solar minimum  and between  5 and 6  per day  in solar maximum, these numbers also depend  on the used coronagraphs \citep{StCyr2000,Webb2012,Lugaz2017}.  They are  originated from highly-sheared magnetic field regions  which can be refereed as  large magnetic flux ropes carrying strong electric currents.} They  are statistically more likely to lead to geomagnetic disturbances when their solar sources are facing the Earth \citep{Bothmer07,Bein11,Wimmer14}. According to their speed, their interplanetary signatures (ICMEs)  may reach the Earth in one to five days after the flare \citep{Yashiro06,Gopalswamy09,Bein11}. 

{  Halo CMEs observed with the  white light SMM coronagraph were  firstly named ''global CMEs''  \citet{Dere2000} and already suspected to be responsible of geoeffective events \citep{Zhang1988}. }Recent studies confirmed the geoeffectivity of halo CMEs  which generally  form magnetic clouds (MC) (e.g. Bocchialini et al 2017, Solar Physics in press). The  MCs are associated with extreme storms ($Dst < -200$ nT) and  intense storms ($-200 <  Dst < -100$ nT) \citep{Gonzalez07,Zhang07}, while the moderate storms  ($-100 <  Dst < -50$ nT) studied in the solar cycle 23 were found to be associated with co-rotating regions  by $47.9 \%$, to ICMEs or magnetic clouds (MC)   by $20.6 \%$, to sheath fields  by $10.8 \%$, or to combinations of sheath and ICME ($10\%$)  \citep{Echer13}.

{ However magnetic clouds can be not so effective if  they are directed away from Earth like the fast ICME of July 2012 \citep{Baker2013} or if the magnetic field of the cloud arrives  close to the magnetosphere with an orientation towards the North  as for the cases of  August 1972 \citep{Tsurutani1992}.  In August 1972 a huge sunspot group McMath region 11976 (see Figure \ref{spot}) crossed the disk and was the site of energetic flares and consequently shocks were detected at 2.2 AU by Pionneer 10 \citep{Smith1976}. The estimated velocity of the ejecta was around 1700 km/s which is nearly the highest transit speed on record.   \citet{Tsurutani2003} estimated its magnetic field to be around 73 nT which is also a huge number.  But the Dst index indicated a recovery phase relatively low like a moderate storm \citep{Tsurutani1992}.  
Nowaday the  {\it in situ } parameters of the solar wind including the interplanetary magnetic field, IMF,
 are monitored  at L1 by the   ACE spacecraft \citep{Chiu1998} magnetic field (MAG experiment)  or similar instruments. They indicate clearly the passage  of the satellite through an ICME or magnetic cloud by the  changes of the solar wind speed, the reversed sign of the magnetic components Bx and By. The ICME is  more geoeffective if the IMF-Bz component is negative indicating a strong coupling with the magnetosphere. }
 
{  We can conclude that   if extreme solar storms do not necessary initiate   extreme geomagnetic events, extreme geomagnetic events are  nearly always produced by extreme solar storms. And extreme solar storms are most of the time  issued from the biggest sunspot groups  which produce the most energetic events \citep{Sammis2000}.}

%Statistical analysis made  in the solar cycle 23 (1996-2008) showed that the interplanetary drivers of the moderate storms  ($-100 <  Dst < -50$ nT) were found to be associated with co-rotating regions  by $47.9 \%$, to ICMEs or magnetic clouds (MC)   ($20.6 \%$), to sheath fields ($10.8 \%$), or to combinations of sheath and ICME ($10\%$)  \citep{Echer13}. Intense  storms ($-200 <  Dst < -100$ nT) are mainly associated  with  MCs  \citep{Gonzalez07,Zhang07}. \\

%A recent analysis of all the solar events occurring in 2002 proved that the importance of getting   solar storms  linked to geoeffectivity is to have a large halo CME expelled from   a solar source near the central meridian, and  followed by  a magnetic cloud in the heliosphere (Bocchialini et al 2017). { A halo CME consists of  bright emission  located all around the Sun. They were first detected in the white light coronagraphs and it has been suggested that these bright emission could be related to intense  geoffective events \citep{Zhang1988}. When the LASCO coronagraph aboard SOHO was launched, they were called  '' global '' CME  \citet{Dere2000}.}

The paper is organized as following. After an historical review of large sunspot groups observed on the Sun related to geomagnetic storms (Section 2),  we present statistical results on  star and sun flares according to  the characteristics of the spots (flux, size)  (Section3). Section 4 is focused on a MHD model (OHM) predicting the capability of the Sun to produce extreme events.  { Finally the conclusion is given in Section 5.}

%Figure 1

%Figure 2 
\begin{figure}
\centering
  \mbox{
 \includegraphics[width=10cm]{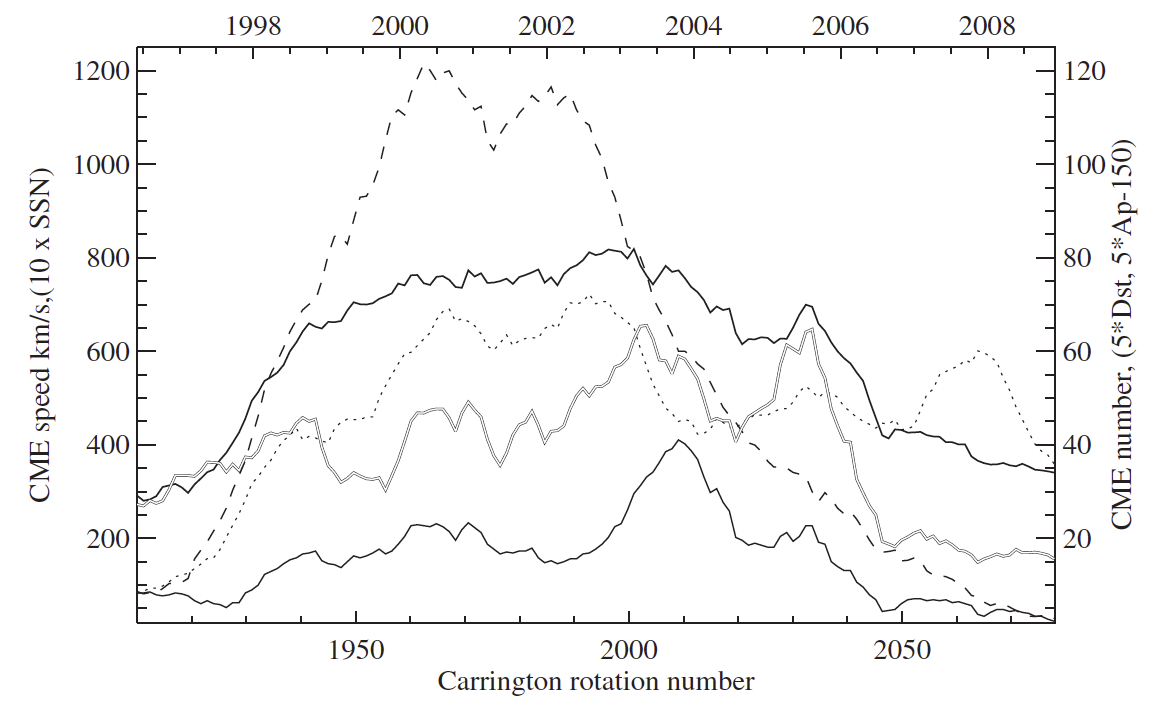}
 }
\caption{{ CME number and speed per solar Carrington rotation related to sunspot number and indexes of geoeffectivity (Dst and Ap).}
 The dashed line shows the sunspot number, the bold solid line the CME
speed index, the dotted line the CME number, the double line the
Dst
index, and the
thin solid line represents the
Ap
index (adapted from \citet{Kilcik11}).}
\label{CME}
\end{figure}

%Figure 3

\begin{figure*}
 \centering
  \mbox{
 \includegraphics[width=14cm]{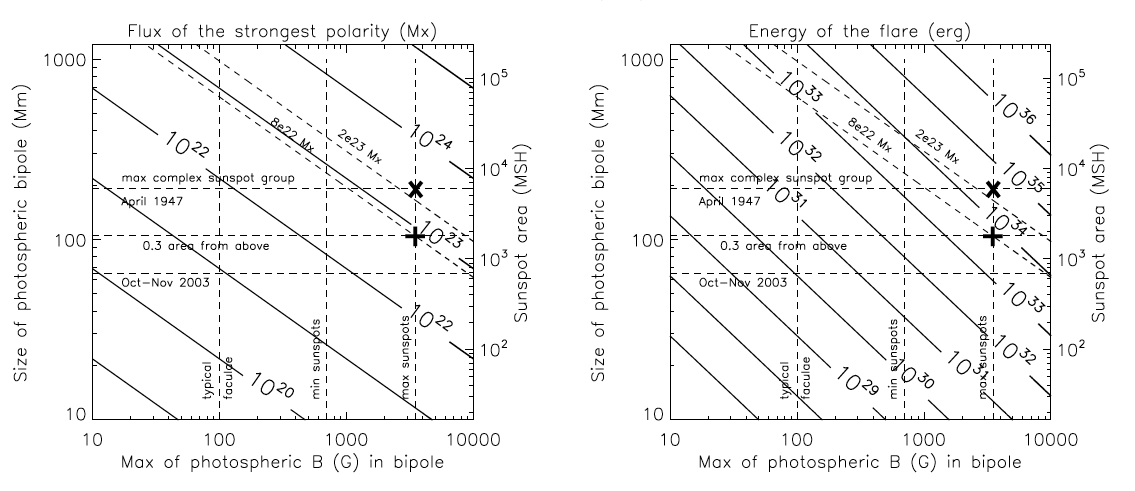}
 }
\caption{
Magnetic flux in the dominant polarity of the bipole, and magnetic energy released during the flare, calculated as a function of the maximum
magnetic field and the size of the photospheric bipole. The x and + signs correspond to extreme solar values. The former is unrealistic and the
latter must be very rare (from \citet{Aulanier13}.)}
\label{OHM}
\end{figure*}
%\cite{Webb2012} reviewed the origins and characteristics of CME and their associated phenomena, near the Sun and in the heliosphere, in terms of space weather.}
%Our innovative approach is based on a statistical analysis of CME's geoeffectiveness during the solar maximum of the cycle 23 (the whole year 2002). This unprecedented study consists in: starting from the list of SSC (Appendix \ref{App:Appendixalldata}, section \ref{section:App_SSC} and Table \ref{table:SSC_list}) and, linking as far as possible each SSC to a CME, halo or not, but with a clear source on the Sun in a temporal window determined by considering two extreme propagation velocities ($300$ and $1\ 500 $ km.s$^{-1}$) \citep{Bein2012}. When there is no CME identified as the source of the observed SSC, we seek what process in the solar wind may lead to the SSC, with the help of observations at L1.

\section{Historical view of solar sources of geoeffectivity}

The Carrington event in September 1, 1859, well known to be one of the largest solar Sunspot groups leading to one of the strongest
flare \citep{Carrington1859,Hodgson1859}   had the largest magnetic signature ever
observed at European latitudes with the  consequent aurora visible at low { geographic} latitude  ($\pm18^\circ$)   observed 17.5 hours later.  Using the transit time, \citet{Tsurutani2003}   proposed that the $Dst$  value decreased down to $-1\ 760$ nT  during this event. { The Colaba  (Bombay) record  allowed to have a more precise determination around -1600 nT  \citep{Cliver2013,Cid13}. This value is  more than twice the value of the next extreme geomagnetic events.}
Revisiting this event by analysing  ice core nitrates and $^{10}Be$ data,
\citet{Cliver2013} claimed that it reached only $-900$ nT.   Nevertheless it { seems}  to be the strongest geoeffective  event registered up to now. A correlation between solar energetic proton fluence (more than $30$MeV) and flare size based on modern data  proves that this event can be classified as an extreme solar event with a X-ray flare having  an estimated  energy  larger than $ X10$.
All these extreme registered events, 12 episodes since the Carrington events, are solar activity dependent  \citep{Gonzalez11a} (rough association).  They occurred mainly during solar cycle maximum of activity with its two bumps and  a secondary peak during the declining phase of the solar cycle. 
% { Between the two peaks of activity there is a dip called ''Gnevyshev gap'' explained by the change of  the longitudinal location of active regions which produces a non-axial magnetic open field \citep{Wang2006}.  The second peak is due to the CMEs which continue to occur with no associated flares. }

Between 1876 and 2007, the largest sunspot area   overlaid by large bright flare ribbons  was observed in the Meudon spectroheliograms in Ca II K1v and  H$\alpha$   between  July  20-26  1946 \citep{Dodson1949}.
A well observed   flare event occurred on  July 25 1946  at 17:32 UT and  caused a huge geomagnetic storm 26.5 hours later. 
The size of the  sunspot was equivalent to 4200 millionths of the solar hemisphere (MSH)  and the ribbon surface around 3570 MSH \citep{Toriumi16}; 
 The Carrington AR  sunspot group seemed to be smaller than that one according to the sunspot  drawings.

The next year an even larger sunspot was visible in  { the  spectroheliogram of } April 5,  1947 with a size reaching 6000 MSH but had no geoeffectivity effect (Figure \ref{spot}). The flare looked to be extended and powerful but 
not accompanied by coronal  mass ejections. It could be a similar case  to the more recently  event observed in October 2014. The AR 12192 presented a sunspot area of 2800 MSH and was the site of several flares (6 X- and 24 M-class) \citep{Sun15,Thalmann15}. These two active regions are really exceptional. The AR 12192 did not launch any CMEs. Different interpretations have been proposed:  the region would possess not enough stress, no enough free energy.  Or the  CME eruptive flux rope would not have reached the threshold height  of the torus instability \citep{Zuccarello15}.

Although there are in average two CMEs per day, only  some of them are geoeffective.  In October and November 2003, the largest sunspot groups  (AR 10486 with an area of 3700 MSH), crossed the disk  and were the sites of extreme events (Figure \ref{spot}). X 17, X 10 and X 35 flares were reported on October 28, October 29  and November 4 respectively. However the more  extreme geomagnetic  storm  occurring  during   the whole Solar Cycle 23 with a $Dst =-422$ nT  was linked to a M9.6 class flare on November 20, 2003  \citep{Gopalswamy05,Moestl08,Marubashi12}. The origin of the solar event was  in the region AR 10501  and has been associated with the eruption of a large  filament  \citep{Chandra10} (Figure \ref{spot}).  
The AR 10501 had  not the largest  sunspot area but the cause of the flare and  CME was merely due to the injection of opposite magnetic helicity by a new emerging flux which produced a destabilization of the large filament and  lead to a  full halo  CME (speed = 690 km/s) and a magnetic cloud in the heliosphere.  The size of the sunspot is an important parameter but it is not sufficient to get an extreme solar storm.

Since the geoeffectivity is not straightforward, in order to forecast major storms, it is important to understand the nature (magnetic strength and helicity) and the location of the solar sources, the propagation of the CMEs through the interplanetary medium and their  impacts on the magnetosphere/ionosphere system. Statistical studies of solar and magnetic activities during solar cycle 23 have permitted to associate CMEs and geomagnetic disturbances,  providing  long lists of CMEs with their characteristics i.e.  their width, velocity, and solar sources \citep{Zhang07,Gopalswamy10a, Gopalswamy10b}. They showed that a CME would more likely give rise to a geoeffective event if its characteristics are: a fast halo CME  (with an apparent width around $360^\circ$) and a solar source close to the solar central meridian. 

In some cases, the proposed sources came from active regions close to  the limb. \citet{Cid12} proposed to revisit this subset of events: in order to associate  every link in the Sun-Earth chain, they have not  only considered  the time window of each CME-ICME, but also they  have carefully revised every candidate at the solar surface. 
%The links between CMEs and geomagnetic events were obtained through$in$-$situ$ measurements (Advanced Composition Explorer - ACE, Wind satellites, surface geomagnetic observatories); those between CMEs and solar sources were obtained by using imagers and coronagraphs (SOHO). 
The result was that a CME coming from a solar source close to the limb cannot be  really geoeffective  (i.e, associated with a at least moderate and a fortiori intense storm) if it does not belong to a complex series of other  events.   {  
Possible deflection of a CME in the corona as well as  in the interplanetary space  may change the geoeffectiveness of a CME \citep{Webb2012}.
 It has been reported   deflection up   a few ten  degrees,  even during the SMM mission \citep{Mein1982,Bosman2012,Kilpua2009,Zuccarello2012,Isavnin2013,Mostl2015}.} 
In the statistical analysis of  Bocchialini et al 2017, it has been shown that a  CME  deflected from its  radial direction  by more than 20 degrees produced  an exceptional geoeffective event.  { Moreover the orientation of the magnetic field of the magnetic cloud  ($Bz <0$) is also an important parameter to get an extreme geoffective event (see the Introduction).}

\section{Characteristics of super flares}
Free magnetic energy stored in the atmosphere  is released through global solar  activity  including CMEs (kinetic energy),  flares and SEPs (thermal and non thermal  energy).
There is not really  a physical reason to have a relationship between the different  categories of  released energy.  
\citet{Emslie12} estimated all  energy components  for 38 solar eruptive flares observed between 2002 and 2006.  The maximum of non potential energy  in an active region reached  3$\times 10^{33}$ erg and therefore could power all  flare activity in the region.  0.5 percent of CMEs have a kinetic energy reaching 3 $\times 10^{32}$ erg, otherwise the mean kinetic energy  of 4133 CMEs is around 5 $\times 10^{29}$ erg. They found a weak relationship between the estimations of the different energies  due to  large uncertainties. However  the relationship looks to be more reliable for extreme events (syndrome of the big flare).
However the  systematic study of geoeffective events occurring  through the solar maximum activity year (2002)  already mentioned in Section 1,  showed that only 2 X-class flares among the 12 X-class flares were related to  Sudden Storm Commencement (SSC)  leaded events in the magnetosphere, the other SSCs  were related to M and even C class flares  (Bocchialini et al 2017).  The solar cycle variation of the {\it Dst } does not follow the  general trend of the sunspot number during the
declining phases of  solar cycles but  is comparable to the 
 trend of CME speeds,  and CME numbers  with the secondary peak \citep{Kilcik11} (Figure \ref{CME}).  This behaviour confirmed the importance of CME in the geoeffectivity.\\

However statistical analysis of flare intensity  showed  a  relationship with some categories of active regions.  Flares were related to large sunspot active regions (category A, B, F ) in the classification of Zurich \citep{Eren17}. The class F  consists of  large   ARs with sunspot  fragmentation, indicating commonly the existence of  strong shear.
This study  confirmed the finding concerning the  historical events that large geoffective effects are  linked  to the existence of large sunspot groups \citep{Carrington1859,Dodson1949}.  { The extreme events should be related to large sunspots  like for the ''Halloween'' events on October-November 2003  in AR 10486 (Figure \ref{spot} top right). The flare on  November 4 2003, is generally considered to be
the most intense SXR event during the space age, with an estimated
peak SXR classification ranging from X25 to X45   \citep{Gopalswamy05,Cliver2013}. However the most geoeffective event occurred on the 20 November   2003. The AR 11501 has not a large sunspot and the  solar extreme event is a  coronal mass ejection with large kinetic energy. This  event shows one example of large geoffectivity  not related to the sunspot size  (Figure \ref{spot} bottom row) but to the magnetic shear and magnetic helicity injection \citep{Chandra10}.}
% even there are some exceptions (AR 11501  November 2003 see Figure \ref{spot}).

%After 2002 October  the average maximum speed began to decline and the trend of the decline is similar to that displayed by theDst index
%and is also consistent with the sunspot numbers. However, in year 2004, both the CME speed andDstindices began to increaseagain and their values peaked during the middle of year 2005(CR 2035), all while the sunspot number continued to vanish. 

Recently super flares (energy  $ 10^{34}$  to  $ 10^{36}$  erg)  have been discovered in Sun-like stars (slow rotating stars)  by the Kepler new  space satellite  \citep{Maehara12}. A debate started about the possibility of observing such super flares on the Sun.  \citet{Shibata13}  forecasted that one such super flare could occur every 800 years. Stars are suspected to have large spots and a large  sunspot  on the Sun with a flux of 2 $\times$ 10$^{23}$ Mx flux  would be  not impossible and would correspond to an energy of 10$^{34}$ erg \citep{Shibata13}.

\citet{Toriumi16} made a statistical analysis of the new solar Cycle 24  flares between May 2010 and April 2016. Considering 51 flares  exhibiting two flare ribbons (20 X and 31 M-class), they  determined an empirical relationship between  the size of sunspots (S$_{spot}$) in flaring active regions and the magnetic flux $\Phi_{spot}$ in logarithm scale.\\
log $\Phi_{spot}$=0.74 $\times$  log S$_{spot}$  +20 with some uncertainties. \\

Considering the largest spots ever observed on the Sun (July 1946 and  October 2014) they extrapolated this relationship and estimated  a maximum flux of 1.5$\times 10^{23}$ Mx.  They did not take into account the fact that all the energy of the spots can be  transformed in thermal and non thermal energy and not in kinetic energy  (no CME was  launched  in October 2014 for example). 

%This second condition was not filled up.  The ratio between the area of flare ribbons normalized by the sunspot area could be a proxy of  of the size of the stressed region concerned by the flare and the distance between the ribbons a proxy of the importance of the CME. Larger ratio and larger distance were found for CME eruptive events.  compared to compacted flare regions  \citep{Toriumi16}. The ratio was found close to 1 for historical events (1946), this conclusion allowed to consider huge sunspots as being an indicator of large amount of stored  magnetic flux. 

%Figure 4

\begin{figure*}
 \centering
  \mbox{
  \includegraphics[width=12cm]{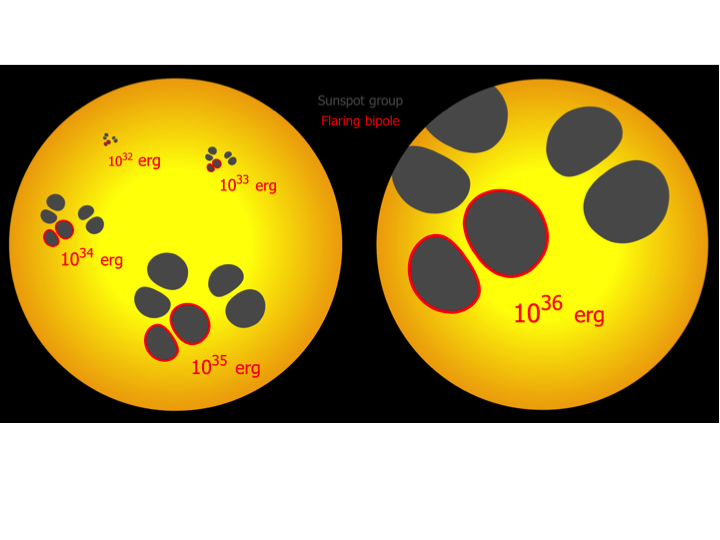}
  }
\caption{Schematic representation of several modeled sunspot groups  without faculae on the solar disk, with their corresponding modeled flare energies computed with the OHM simulation. {  A sunspot group consists of several pairs of sunspots. In each group a   pair of sunspots (surrounded by red curve) representing  1/3 of the  sunspot group area,   is modeled in the simulation. The size of the  grey areas  is normalized to  the size of the spots considered in the simulation (adapted from \citet{Aulanier13}).}}
\label{star}
\end{figure*}

\section{Prediction of extreme solar storms}

It appears that MHD simulations  of emerging flux could be  used to have a systematic survey  to investigate the process of energy storage and find the relationship between sunspot size,  CME eruptive events. 
The  {\it Observationally  driven High order scheme Magnetohydrodynamic code} (OHM) \citep{Aulanier05,Aulanier2010}  simulation has been used as a tool  to experiment huge energetic events on the Sun  e.g. large super flare ($10^{36}$ erg) by varying the characteristics of the sunspots in a large 
  parameter space \citep{Aulanier13}. 
  %The OHM model has been used for such experiment to define the conditions to get \citep{Aulanier2010,Aulanier13}.
   The model consisted of a bipole with two rotating sunspots which is equivalent to create along the polarity inversion line a strong shear with cancelling flux. The    3D    numerical    simulation    solved the full MHD equations for the mass density, the fluid velocity u, and the magnetic field B under the plasma $\beta$ =0 assumption. The calculations were performed in non-dimensionalized
units, using $\mu$ = 1.
  %   $\beta$ =0  but two forces are operand: the magnetic pressure and the gravity. 
  The magnetic field diffusion  favored the expulsion of the flux rope. The space parameter study lead  to graphs of values of  magnetic flux and energy according to the size of sunspot in  MSH  units and the stress of the field (Figure \ref{OHM}). 

The magnetic flux  $\Phi$ and the total flare energy  E are defined as following:\\
%$\Phi$ =42 (B$_z$ /8T) (L$^{bipole}$/5m)$^2$ Wb\\
%E= 40 / $\mu$ (B$_z$ /8T)$^2$ (L$^{bipole}$/5m)$^3$ J \\

\noindent  $\phi$ = 42 $(\frac{B_z}{8T})$  $(\frac{L^{bipole}}{5m})^2$ Wb   \\

\noindent E= $\frac{40}{\mu(\frac{B_z}{8T})^2(\frac{L^{bipole}}{5m})^3 }$J
\\
\\
B is the strength of the magnetic field in the bipole (sunspot), L is the size of the bipole. 
The problem is the estimation of the value  L.
L$^2$ can be computed as the area of an active region with facula (L=200 Mm), The maximum value for the flux is   $\phi$ = 10$^{23}$ Mx and for the energy  E =3 $\times$ 10$^{34}$ erg  that  falls in the range of star superflares \citep{Maehara12}.  However  L should  be reduced to 1/3 due to the fact that the stress of the field concerned only a small part of the PIL \citep{Aulanier13}.   The maximum of energy could not exceed 10$^{34}$ erg. These results come from a self  consistent model  with shear flux leading to  CME with no approximation.
On the other hand the estimations of \citet{Toriumi16}  are very empirical  mixing  different observations not related one to the other one. Each estimation has been overestimated. For example the volume of the active region concerned by the flare has been estimated by the product  of S$_{ribbon}$ (surface area of the ribbons) and distance between the ribbons \citep{Toriumi16}.  However  the uncertainty  on the estimation of the magnetic field in this  volume can lead to an  overestimation by one to two orders of magnitude { according to the  f   value introduced in their  equations}.  Taking unrealistic values  of B and  flux  lead to  unrealistic energy  values never observed in our era  \citep{Emslie12}.  

\section{Conclusion}

{ Commonly extreme solar events are produced in active regions having a strong magnetic reservoir (high magnetic field  and stress). There are defined as very powerful X ray flares, coronal mass ejections with high kinetic energy faced to the Earth leading to magnetic cloud arriving at the magnetosphere with a good orientation (B$_z$ negative) and strong ejections of energetic particles  (SEPs). Large sunspot groups with fragmentation are good candidates for extreme solar storms \citep{Sammis2000}.}

With our Sun as it is today, it seems impossible to get larger sunspots and super-flares with energy $>$ 10$^{34}$ erg. {  Figure \ref{star}   shows 
different sunspot groups. In each of them  a pair of sunspot  surrounded by red curves represents  the bipole used as boundary condition of the OHM  simulation. The energy mentioned below the  pair is the result of the simulation. With huge sunspots   we  obtain   large energies as it is  recorded   for stars  by the  Kepler satellite. Such large spots
have  never been observed on the Sun.}
 We should not forget that the simulation concerns a bipole with rotating spots imposing a strong shear along the PIL. The shear  is  a necessary ingredient to have expulsions of CMEs in the simulation and also in the observations. In order to produce stronger  flares the Sun-like stars  should  have  a much stronger dynamo than the Sun and  a rotation rate exceeding several days. The prediction of having  extreme solar storms in 800 years would be very speculative.

%Here are two sample references: \cite{Feynman1963118,Dirac1953888}.

%\section*{References}
Acknowledgements\\
The author would like to thank  the organizers of the meeting Drs.  Katya Georgieva  and  Kazuo  Shiokawa to invite me  in Varna for the VarSITI meeting in June 2016. I want to thank G.Aulanier for his fruitful comments on this work.
%\end{acknowledgements}

%\bibliography{bibliography}

\end{document}